\documentclass[%
 reprint,
 amsmath,amssymb,
 aps,superscriptaddress
]{revtex4-2}

\usepackage{graphicx}% Include figure files
\usepackage{dcolumn}% Align table columns on decimal point
\usepackage{bm}% bold math
\begin{document}

\preprint{APS/123-QED}

\title{Localized character of charge excitations for La$_{\bm{2-x}}$Sr$_{\bm x}$NiO$_{\bm{4+\delta}}$\\ revealed by oxygen $\bm K$-edge resonant inelastic X-ray scattering}% Force line breaks with \\
\thanks{A footnote to the article title}%

\author{Kohei Yamagami}
\affiliation{Institute for Solid State Physics, The University of Tokyo, Kashiwa 277-8581, Japan}
\author{Kenji Ishii}
\affiliation{Synchrotron Radiation Research Center, National Institutes for Quantum and Radiological Science and Technology, Hyogo 679-5148, Japan}
\author{Yasuyuki Hirata}
\author{Keisuke Ikeda}
\author{Jun Miyawaki}
\author{Yoshihisa Harada}
\affiliation{Institute for Solid State Physics, The University of Tokyo, Kashiwa 277-8581, Japan}
\author{Masanori Miyazaki}
\affiliation{Graduate School of Engineering, Muroran Institute of Technology, 27-1 Mizumoto, Muroran 050-8585, Japan}
\author{Shun Asano}
\author{Masaki Fujita}
\affiliation{Institute for Materials Research, Tohoku University, Sendai, 980-8577, Japan}
\author{Hiroki Wadati}
\affiliation{Institute for Solid State Physics, The University of Tokyo, Kashiwa 277-8581, Japan}
\affiliation{Graduate School of Material Science, University of Hyogo, Ako, Hyogo 678-1297, Japan}

\begin{abstract}
We performed a resonant inelastic X-ray scattering (RIXS) study of La$_{2-x}$Sr$_{x}$NiO$_{4+\delta}$ (LSNO) at the oxygen $K$ edge to investigate the nature of the doped holes with regard to charge excitations.
Charge excitations of the hole-doped nickelates are found to be almost independent of momentum transfer, indicating that the doped holes are strongly localized in character.
Additionally, conspicuous changes in energy position are in temperature dependence.
These characters are observed in stark contrast to those of the high-$T_{c}$ cuprate La$_{2-x}$Sr$_{x}$CuO$_{4}$ (LSCO), where delocalized doped holes form charge excitations with sizable momentum dependence in the CuO$_2$ plane.
This distinct nature of charge excitations of doped holes is consistent with the metallicity of the materials and could be caused by strong electron-phonon coupling and weak quantum spin fluctuation in the nickelates. 
\end{abstract}

\maketitle

%\tableofcontents

Effects of carrier doping are a central issue in strongly correlated electron systems~\cite{TMOreview}.
Carrier doping into insulating transition-metal oxides causes a variety of interesting and important physical properties, such as high-$T_{c}$ superconductivity, a colossal magnetoresistive effect, and metal--insulator transitions.
The parent insulators are categorized as either Mott insulators or charge-transfer insulators~\cite{MottCTGap}.
In the former, on-site Coulomb repulsion of the $d$ electrons ($U$) is less than the energy of the charge transfer ($\Delta$) from the oxygen 2$p$ to the transition-metal $d$ levels.
In the latter, however, $U$ is greater than $\Delta$; this is the case for the layered perovskite lanthanum cuprate La$_{2}$CuO$_{4}$ (LCO)~\cite{LSCOXPSI,LSCOXPSII,LSCOXPSIII,LSCOXPSIV} as well as its isostructural nickelate La$_{2}$NiO$_{4+\delta}$ (LNO)~\cite{LSNOXPS}.
These parent materials are quasi-two-dimensional Heisenberg antiferromagnets with a weak inter-layer coupling of $S = 1$ (Ni$^{2+}$: $d^{8}$) for LNO and $S=1/2$ (Cu$^{2+}$: $d^{9}$) for LCO and show antiferromagnetic order at 330 K and 270 K, respectively~\cite{LNONeel,LCONeel}.
When holes are doped into LNO and LCO by substitution of Sr$^{2+}$ ions for La$^{3+}$ ions, they predominantly occupy the oxygen 2$p$ orbitals.
Despite these similarities, the physical properties of cuprates and nickelates differ considerably.
For instance, whereas a small amount of hole doping ($x\sim0.05$) induces a metallic state and superconductivity in La$_{2-x}$Sr$_{x}$CuO$_{4}$ (LSCO)~\cite{LSCOResistivity}, La$_{2-x}$Sr$_{x}$NiO$_{4+\delta}$ (LSNO) remains insulating up to $x=0.8$~\cite{LSNOResistivity}.

Recently, charge ordering in cuprates has attracted great interest as a competing phase of superconductivity.
Originally, a charge--spin stripe ordering propagating along the Cu-O-Cu bond in the La-based cuprates was discovered, associated with suppression of the superconducting transition temperature near the hole density of $1/8$~\cite{LSCOCOI}, and it is now established that the charge order appears universally in carrier-doped superconducting cuprates~\cite{COreview}.
In the case of LSCO, the charge order was reported for $x=0.11-0.13$~\cite{LSCOCOII}.
The charge--spin stripe ordering also exists in LSNO.
However, it extends over a much wider hole concentration range ($0.22\le n_{h}\le0.50$, $n_{h}=x+2\delta$)~\cite{NeuCOI,NeuCOII,EleCO,XraCOI,XraCOII} than in LSCO, and the transition temperatures of the charge and spin order ($T_{\rm CO}$ and $T_{\rm SO}$, respectively) reach their maximum at $n_{h}=0.33$~\cite{TCOTSO}.
Additionally, the nickelate exhibits an ordered state with diagonal stripes propagating along the diagonal of the NiO$_{2}$ square lattice.
These contrasting physical properties of the hole-doped compounds indicate that the nature of the doped holes differs between LSCO and LSNO.

The existence of oxygen 2$p$ holes in the doped charge-transfer insulators has been confirmed by X-ray absorption spectroscopy (XAS) at the oxygen $K$ edge~\cite{LSCOoxyXASI,LSCOoxyXASII,LSNOoxyXASI,LSNOoxyXASII}.
A pre-edge peak corresponding to the transition of 1$s$ electrons to 2$p$ holes emerges in the XAS spectra, and its intensity increases with hole concentration.
Resonant inelastic X-ray scattering (RIXS) is a natural extension of XAS for the study of momentum-resolved excitations of the 2$p$ holes.
However, most RIXS studies at the oxygen $K$ edge performed to date have focused on undoped compounds and discussed charge-transfer, double-spin-flip (two-magnon), and phonon excitations~\cite{OKedgeRIXSThe,OKedgeRIXSI,NiORIXS,OKedgeRIXSII,OKedgeRIXSIII,OKedgeRIXSIV,OKedgeRIXSV,LSCOoxyRIXS,OKedgeRIXSVI,OKedgeRIXSVII}.
Alternatively, charge excitations in the doped compounds were studied by RIXS at the transition-metal $K$ edge~\cite{LSCOKRIXSI,LSCOKRIXSII,NiKedgeRIXSI,NiKedgeRIXSII,NiKedgeRIXStemp}.
Although charge excitations below the charge-transfer gap are observed, their intensity is very weak.
In particular, low-energy excitations (below several hundred millielectronvolts) are difficult to resolve from the huge tail of the elastic scattering.
The difficulty lies mainly in the choice of absorption edge; namely, the sensitivity of the transition-metal $K$ edge is low, and O $K$-edge RIXS is more suitable for studying the excitations of doped holes with strong O 2$p$ character.
A recent O $K$-edge RIXS study of LSCO revealed momentum-dependent charge excitations; the magnitude of the dependence is on the order of the hopping energy ($\sim$0.4 eV)~\cite{LSCOoxyRIXS}.
We note that electrons are doped in the Cu 3$d$ orbital in the cuprates, and charge excitations in the electron-doped cuprates are clearly observed in the RIXS at the Cu $K$ and $L_{3}$ edges~\cite{NCCOKRIXSI,NCCOKRIXSII,NCCOLRIXSI,NCCOLRIXSII,NCCOLRIXSIII}.
It is therefore appropriate to extend O $K$-edge RIXS to another hole-doped charge-transfer insulator, LSNO.

In this paper, we report O $K$-edge RIXS measurements of LSNO to investigate charge excitations of the doped holes.
Observed charge excitations are strongly localized in character owing to their very weak momentum dependence in O $K$-edge RIXS spectra.
Furthermore, a change in the position of the low-energy charge excitation energy at several hundred millielectronvolts is observed at the charge-stripe ordering transition temperature.
We discuss the possible origin of localized character of the charge excitations.

\begin{figure}
\begin{center}
\includegraphics[width=8.3cm]{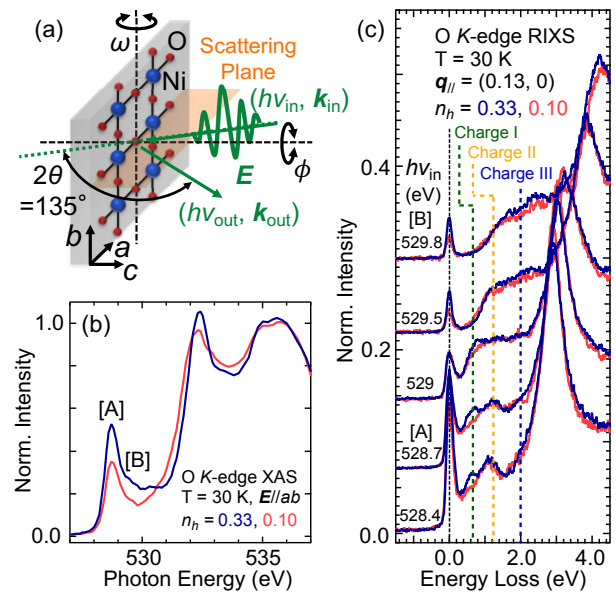}
\end{center}
\vspace{-8mm}
\begin{center}
\caption{(Color online) (a) Geometry of O $K$-edge RIXS measurement experiments. ($h\nu_{\rm in}$, ${\bm k}_{\rm in}$) and ($h\nu_{\rm out}$, ${\bm k}_{\rm out}$) denote the photon energy and the momentum vector of incident and scattered X-rays, respectively. (b) $n_{h}$ dependence of O $K$-edge XAS spectra for LSNO with ${\bm E}//ab$ in PFY mode at $T=30$ K. (c) $h\nu_{\rm in}$ dependence of O $K$-edge RIXS spectra at $\bm q_{//}$ = (0.13, 0). $h\nu_{\rm in}$ values of 528.7 eV and 529.8 eV correspond to the photon energy of peaks [A] and [B], respectively, in O $K$-edge XAS spectra. The vertical dashed lines at 0.6 eV, 1.2 eV, and 2.0 eV indicate the peak positions of the three charge excitations (Charge I, II, and III) discussed in the text.}
\label{Fig.1}
\end{center}
\end{figure}

The O $K$-edge XAS and RIXS measurements were performed at the HORNET end-station of the SPring-8 BL07LSU beamline~\cite{HORNET,BL07LSU}.
Single crystals of LSNO with ($x,\delta,n_{h}$) = ($0.00,0.05,0.10$) and ($0.33,0.00,0.33$), prepared by the traveling-solvent floating-zone method, were cleaved parallel to the crystalline $ab$-plane in the air before the measurements were taken.
XAS spectra in the partial fluorescence yield (PFY) mode were recorded using a silicon drift detector.
A linear background evaluated at 520--526 eV was subtracted, and normalizations were carried out using the intensity at 536 eV, corresponding to the main peak of O 1$s$ $\rightarrow$ 2$p$ absorption~\cite{RSNOoxyXAS}.
All O $K$-edge RIXS spectra were obtained by subtracting a constant background and normalizing by the total RIXS intensity.
The energy resolution of RIXS was set to $\sim$140 meV, determined by the elastic peak for a gold plate.
XAS and RIXS measurements were carried out at a range of temperatures from 30 K to 350 K.
The experimental configuration is depicted in Fig.~\ref{Fig.1}(a).
The $\sigma$-polarized X-ray having an electric field vector ($\bm E$) perpendicular to the scattering plane was irradiated on the $ab$-plane (${\bm E}//ab$), where the scattering plane corresponds to the plane defined by two wavevectors of the incident (${\bm k}_{\rm in}$) and scattered (${\bm k}_{\rm out}$) X-rays.
The scattering angle (2$\theta$) was fixed at 135$^{\circ}$, and the crystalline $c$-axis was kept parallel to the scattering plane.
The momentum transfer (${\bm q}={\bm k}_{\rm in}-{\bm k}_{\rm out}$) was tuned by changing the sample ($\omega$) and azimuthal ($\phi$) angles, which control the magnitude of the projected momentum ($\bm q_{//}$) and its direction within the $ab$-plane.
In our measurements, the [$h, 0$] and [$h, h$] directions were set by fixing $\phi$ at 0$^{\circ}$ and 45$^{\circ}$.
The sign of $\bm q_{//}$ is positive when $\omega$ is toward the grazing exit condition.
The $\bm q_{//}$ values were calculated from the magnitude of $\omega$ and $|{\bm k}_{\rm in}|$ and expressed in units of $2\pi/a$ using the tetragonal lattice parameters $a$ = $b$ = 3.82 \AA~\cite{EleCO,LSNOlatticeII,LSNOlatticeIII}.

Figure~\ref{Fig.1}(b) shows O $K$-edge XAS spectra with ${\bm E}//ab$ for LSNO with $n_{h}$ = 0.10 and 0.33.
The pre-edge peaks at 528.7 eV and 529.8 eV, labeled [A] and [B], are observed.
These two peaks are explained by the ligand field theory on a NiO$_{6}$ cluster and assigned as the high-spin ($S=1$) and low-spin ($S=0$) states of Ni$^{2+}$ ions, respectively~\cite{LSNOoxyXASII}.
On the other hand, the spectral feature at 532.5 eV corresponds to the transition to the upper Hubbard band of Ni 3$d_{x^{2}-y^{2}}$ hybridized with O 2$p_{x,y}$ orbitals, since dipole selection rules allow only 1$s$ $\rightarrow$ 2$p_{x,y}$ transitions for ${\bm E}//ab$.
We confirmed the numbers for $n_{h}$ from the integration of the spectral intensity from 525 eV to 530.9 eV ($\mu(n_{h})$) and obtained the value $\mu(0.10)/\mu(0.33)=0.74$, which is consistent with that in the previous work~\cite{LSNOoxyXASII}.

In order to investigate the excitations of the doped holes, the incident photon energy ($h\nu_{\rm in}$) dependence of O $K$-edge RIXS spectra were measured near the pre-edge peaks, from 528.4 eV to 529.8 eV.
Figure~\ref{Fig.1}(c) shows $h\nu_{\rm in}$-dependent O $K$-edge RIXS spectra of LSNO with $n_{h}$ = 0.10 and 0.33.
Whereas the high-energy peak above 3 eV is assigned as a fluorescence component due to the continuous shift to higher-energy sides with increasing $h\nu_{\rm in}$, the excitations at 0.6 eV, 1.2 eV, and 2.0 eV exhibit Raman-like behavior, as energy loss ($h\nu_{\rm in}-h\nu_{\rm out}$) hardly changes with $h\nu_{\rm in}$.
We now discuss the origin of the Raman-like components.
Because the charge transfer gap in the parent LNO is observed at $\sim$4 eV~\cite{LSNOXPS,NiKedgeRIXSI}, energy of the Raman-like components is too low to assign them as interband excitations across the charge-transfer gap.
Spin-flip magnetic excitations are forbidden in the O $K$-edge RIXS and two-magnon excitations are unlikely as an origin of the Raman-like components judging from the one-order smaller nearest neighbor exchange interaction $J \sim 30$ meV \cite{LNONeel, NiLedgeRIXS}.
In fact, two-magnon excitations are observed at $\sim$0.2 eV or below in optical Raman scattering~\cite{LNORaman,LSNORaman}.
Plasmon excitations are also unlikely in the insulating LSNO.
Thus, we consider that the Raman-like components originate from the intraband charge excitations of doped holes.
According to the O $K$-edge RIXS spectra around $h\nu_{\rm in}$ = 528.7 eV, the intensity of the excitation at 0.6 eV clearly evolves with increasing $n_{h}$, whereas the changes in intensity of the excitations at 1.2 eV and 2.0 eV are small. 
This $n_{h}$ dependence suggests that the excitation at 0.6 eV includes doped holes that are by nature stronger than any other excitations.

In optical conductivity of hole-doped nickelates, intraband charge excitations emerge below the charge-transfer gap and its intensity increases with hole density \cite{LSNOopticalI,LSNOopticalV,LSNOopticalII,LSNOopticalIII}.
The conductivity spectra are analyzed theoretically by the numerically exact diagonalization based on the Hubbard model for two-dimensional doped nickelate~\cite{Opticaltheory}.
In this model, the Hamiltonian includes the electron hopping ($\sim$0.3 eV), the intra- and inter-orbital Coulomb interactions, and the Hund's-rule coupling ($\sim$1.4 eV) between electrons in $e_{g}$ orbitals under octahedral symmetry, whereas crystal-field splitting between $e_{g}$ orbitals, O 2$p$ orbitals, and the electron--phonon interaction are neglected.
This calculation demonstrates that motion of doped holes causes two types of charge excitations depending on the spin state of the Ni$^{2+}$ site.
One keeps the $S=1$ state and the other is accompanied with a spin-flip to $S=0$.
The former is characterized by the $^{3}A_{2}$ state of the Ni$^{2+}$ site and the latter has two $d$-electron configurations of the $^{1}E$ and $^{1}A_{1}$ states.
The former excitations appear at lower energy than the latter and the calculation qualitatively explains the multiple peaks in optical conductivity.
Comparing our RIXS spectra with the optical conductivity, the excitation energy of the Raman-like components is found to be roughly the same as that in the optical conductivity.
Therefore, we adopt the assignment of the calculation; the Raman-like component at 0.6 eV is ascribed to the non-spin-flip charge excitations and  the components at 1.2 eV and 2.0 eV correspond to the charge excitations accompanied with the spin-flip at the Ni$^{2+}$ sites.
\begin{figure}
\includegraphics[width=8.5cm]{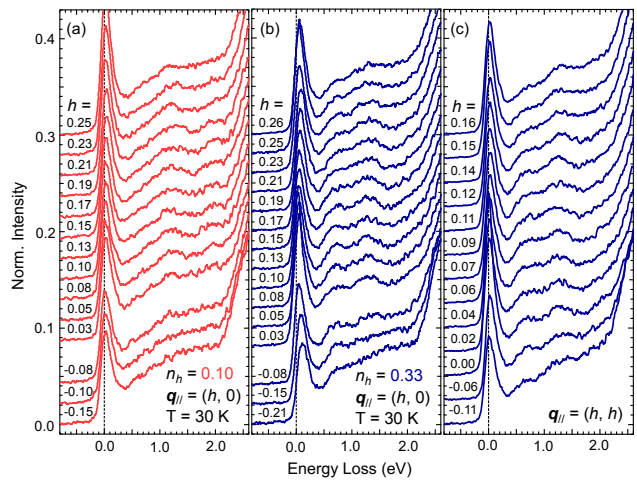}
\vspace{-8mm}
\begin{center}
\caption{(Color online) $\bm q_{//}$ dependence of O $K$-edge RIXS spectra at $T=30$ K for LSNO with (a) $n_{h} = 0.10$ at $(h, 0)$ and $n_{h} = 0.33$ at (b) $(h, 0)$ and (c) $(h, h)$.}
\label{Fig.2}
\end{center}
\end{figure}
\begin{figure*}
\includegraphics[width=17cm]{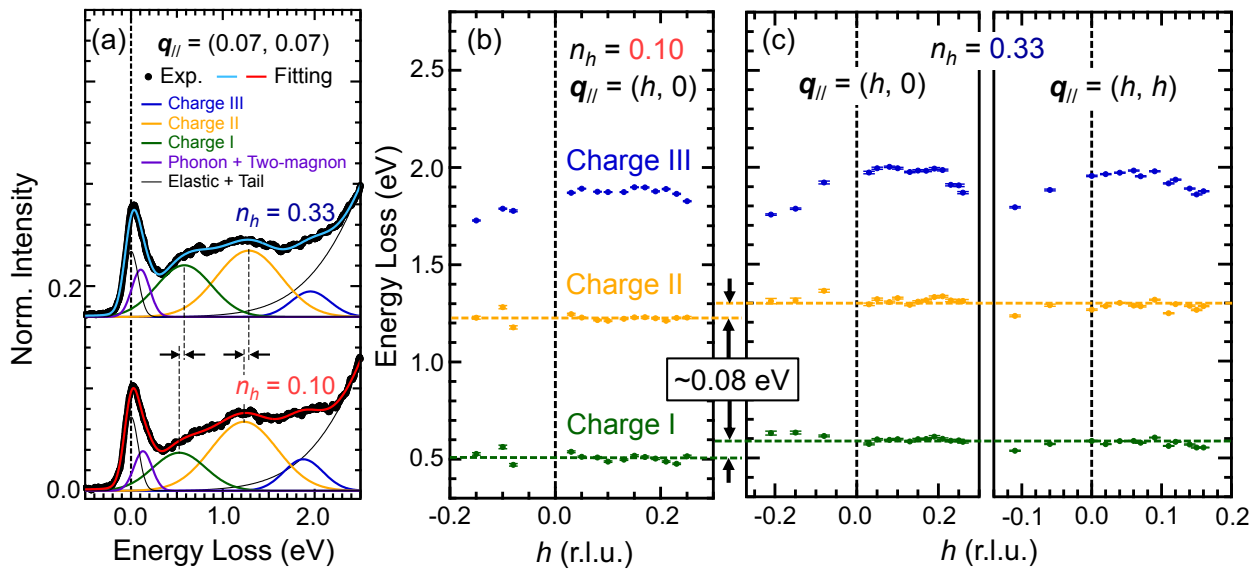}
\vspace{-5mm}
\begin{center}
\caption{(Color online) (a) Comparison between experimental RIXS spectra (dots) and fitting results using multiple Gaussians (lines) for $n_{h}$ = 0.10 and $n_{h}$ = 0.33. (b-c) Momentum dependence of the peak positions of three charge excitations (Charge I, II, and III) obtained from the fitting. The horizontal dashed lines with colors show the averaged value of Charge I and II. The error bars for the peak energy are estimated from the fitting results.}
\label{Fig.3}
\end{center}
\end{figure*}

Next, let us discuss the $n_{h}$ and $\bm q_{//}$ dependence of the charge excitations at 0.6 eV, 1.2 eV, and 2.0 eV at $h\nu_{\rm in}$ = 528.7 eV.
We refer to the excitations as Charge I, Charge II, and Charge III, respectively.
Figures~\ref{Fig.2}(a)--(c) show the $n_{h}$- and $\bm q_{//}$-dependent O $K$-edge RIXS spectra for LSNO at $T$ = 30 K.
Few spectral changes with $\bm q_{//}$(= ($h, 0$) and ($h, h$)) are observed, suggesting that momentum-dependence of the charge excitations is very week.
In order to evaluate the peak position of the charge excitations, we fitted the spectra using multiple Gaussians which model resolution-limited elastic scattering, low-energy peak around 0.1 eV for phonons and two-magnons, three charge peaks (Charge I, II, and III) and the high-energy tail.
Except for the peak of Charge I, we fixed the width of the Gaussians.
Detailed procedure and summary of the fitting analysis are presented in the supplemental materials~\cite{Supplemental}.
As shown in Fig.~\ref{Fig.3}(a), multiple Gaussians can reproduce the experimental spectra very well.
For Charge I and II, the peak position of $n_{h}=0.33$ is found to be higher energy loss side than that of $n_{h}=0.10$.
This energy shift is also observed at other $\bm q_{//}$ values in Fig.~\ref{Fig.3}(b) and (c), where the peak positions obtained from the fitting analysis are plotted as a function of $\bm q_{//}$.
The averaged peak position of Charge I and II for $n_{h}=0.33$ is higher by $\sim$0.08 eV than that for $n_{h}=0.10$.
These figures also confirm that the dispersion width ($W$) of Charge I and II is less than 0.1 eV, and therefore we conclude that these charge excitations of LSNO have a localized character.

At $n_h=0.10$ and $0.33$, doped hole has the $x^2-y^2$ symmetry~\cite{LSNOoxyXASII,RSNOoxyXAS} which is a hybridized state of the O $2p_{x,y}$ and the Ni $3d_{x^2-y^2}$ orbitals.
An electron transfers from O $1s$ state to the hole state at the XAS peak labeled [A] in Fig.~\ref{Fig.1}(b).
Since energy of incident photon, which is polarized parallel to the NiO$_{2}$ plane [Fig.~\ref{Fig.1}(a)], is tuned to the peak [A] for RIXS measurement, the charge excitations observed in this condition are associated with the hole state.
This experimental condition is the same as in the work on LSCO in Ref.~\cite{LSCOoxyRIXS}.
The O $2p_{x,y}$ orbitals hybridized with the Cu $3d_{x^2-y^2}$ orbital get involved in the observed charge excitations in LSCO, namely, charge excitations in the present study of the nickelates have the same orbital character as those of the cuprates in Ref.~\cite{LSCOoxyRIXS}.
Our results confirm that hole dynamics within the plane is significantly different between LSNO and LSCO.
One difference is the multiple component of the charge excitations in LSNO.
This comes from the fact that motion of holes can induce a spin-flip process at the Ni$^{2+}$ site as mentioned above.
Meanwhile, charge excitation in hole-doped cuprates has a single component because the $S=1/2$ state of the Cu$^{2+}$ sites keeps unchanged.
More importantly, we find another difference in momentum scan.
Momentum dependence of the charge excitations of doped holes is almost negligible ($W\le$ 0.1 eV) in LSNO with $n_{h}=0.33$, whereas that in LSCO is remarkable ($W\sim$ 0.5 eV~\cite{LSCOoxyRIXS}).
Notably, this observation agrees with the metallicity of the materials, that is, LSNO is insulating and LSCO is metallic.
The distinct momentum dependence is reminiscent of a study of the transition-metal $K$-edge RIXS~\cite{NiKedgeRIXSI}, where interband excitations across the charge-transfer gap are compared in the parent compounds.
Similarly to the intraband excitations in our study, the interband excitations of the parent LNO do not exhibit momentum dependence in contrast to the sizable dispersion in LCO.
The interband excitation is a superposition of a hole in the valence band and an electron in the conduction band.
Our result confirms that the valence band alone has weak momentum dependence and holes in the band are localized in LSNO.

\begin{figure*}
\includegraphics[width=15cm]{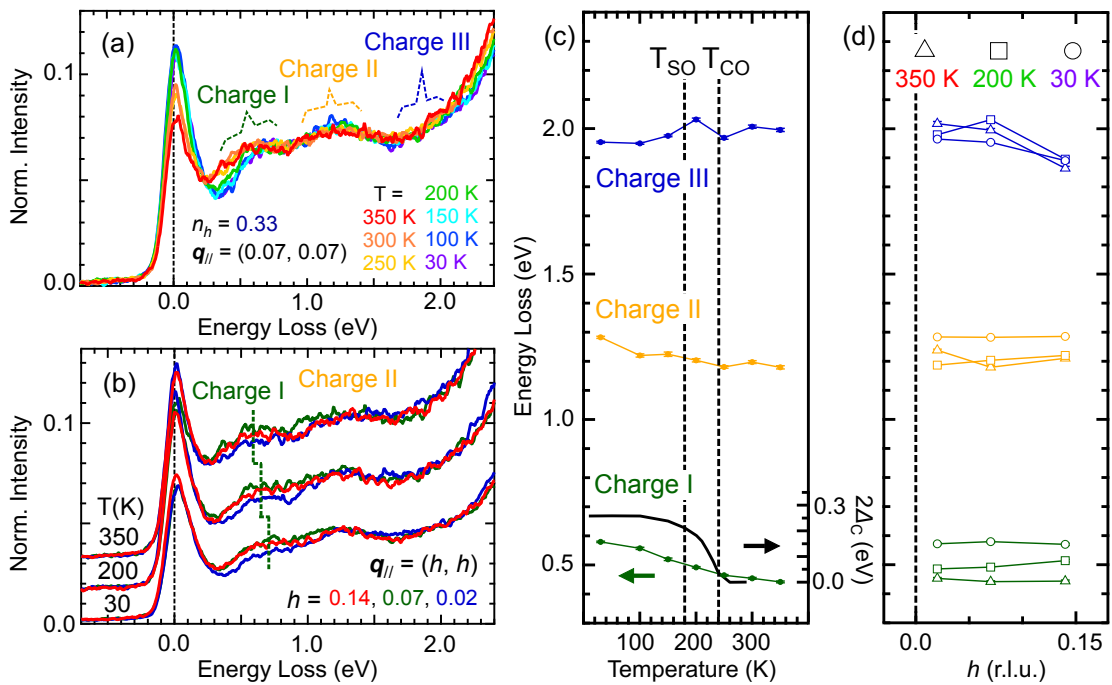}
\vspace{-5mm}
\begin{center}
\caption{(Color online) (a) Temperature dependence of O $K$-edge RIXS spectra for LSNO with $n_{h}$ = 0.33 at $\bm q_{//}=(0.07, 0.07)$. (b) $\bm q_{//}=(h, h)$ dependence of O $K$-edge RIXS spectra at each temperature. (c) Peak positions of three charge excitations obtained by multiple-Gaussian fitting as a function of the temperature. The momentum is $\bm q_{//}=(0.07,0.07)$. Charge gap ($2\Delta_{\rm C}$) reported in Ref.~\cite{LSNOopticalII} is also shown. (d) Momentum dependence of peak positions of three charge excitations at $\bm q_{//}=(h, h)$. The error bars for the peak energy are estimated from the fitting results.}
\label{Fig.4}
\end{center}
\end{figure*}
In the following, we discuss possible explanations for the localized character of the charge excitations of LSNO with $n_{h}=0.33$.
One is the formation of polarons originating from the electron--phonon (EP) coupling in doped holes.
An optical study suggests that the size of polarons in LSNO is smaller than those in LSCO~\cite{EPcouplingI}.
The smaller polaron size would induce localization of charge excitations, and the localized charge excitations in our O $K$-edge RIXS imply a small size for the polarons in LSNO.
Theoretical studies reveal how small polarons could form as a result of EP coupling~\cite{EPcouplingII,EPcouplingIII}.
A four-band model taking Ni $3d_{x^{2}-y^{2}}, 3d_{3z^{2}-r^{2}}$ and O 2$p_{x,y}$ orbitals into account predicts that polarons become stable if $\Delta\simeq U$~\cite{EPcouplingII} and could exist more stably in LSNO than in LSCO because $\Delta\le U$ for LSNO~\cite{LSNOXPS} and $\Delta\ll U$ for LSCO, as obtained by X-ray photoemission spectroscopy~\cite{LSCOXPSI,LSCOXPSII,LSCOXPSIII,LSCOXPSIV}.
Meanwhile, local-density approximation taking the Coulomb interaction into account explains that the large binding energy of polarons comes from an interplay with magnetic confinement to doped holes~\cite{EPcouplingIII}.
Thus, polarons could become smaller as a result of both electronic and magnetic interactions.
Another possible origin is the magnitude of the spin in parent materials.
The quantum spin fluctuation of the antiferromagnetic states in LNO ($S=1$) is much weaker than that in LCO ($S=1/2$).
$^{139}$La nuclear quadrupole resonance and photoemission studies indicate that the spin of the doped hole nearest the Fermi level is antiparallel to the Ni$^{2+}$ spins~\cite{LSNOXPS,LSNONQR}.
This antiferromagnetic coupling between the $d_{x^{2}-y^{2}}$-symmetry hole spin and the Ni spin induces energetically unfavorable hopping of the hole to the neighboring site, which would lead to lower mobility of the doped holes in LSNO with $n_{h}=0.33$.
These localized character of the charge excitations is consistent with the insulating electronic state, thereby lacking a prerequisite of superconductivity, and we speculate that it could cause a wider range of hole densities of charge ordered phases in nickelates than in cuprates~\cite{NSNOCO}.

As mentioned earlier, LSNO with $n_{h}$ = 0.33 has charge--spin stripe ordering, and it is here that the ordering transition temperatures reach their respective maxima ($T_{\rm CO}$ = 240 K, $T_{\rm SO}$ = 180 K)~\cite{TCOTSO}.
We now focus on the behaviors of the charge excitations of doped holes for LSNO with $n_{h}$ = 0.33 across $T_{\rm CO}$ and $T_{\rm SO}$.
Figure~\ref{Fig.4}(a) shows the temperature-dependent O $K$-edge RIXS spectra of LSNO with $n_{h}$ = 0.33 at $\bm q_{//}$ = (0.07, 0.07).
With increasing temperature, spectral weight of Charge I shifts to lower energy together with the broadening of the width of Charge II.
The shift is prominent between 200 K and 250 K.
Similar variations with temperature are observed at other momenta as shown in Fig.~\ref{Fig.4}(b).
We fitted the spectra using multiple Gaussian model~\cite{Supplemental}.
Here, we assume that the width of Charge I and II is the same for all ${\bm q_{//}}$ but it varies with temperature.
Peak positions obtained from the fitting analysis are shown in Fig.~\ref{Fig.4}(c) and (d).
Peak energy of Charge I is shifted to lower energy by $\sim$0.14 eV with increasing temperature [Fig.~\ref{Fig.4}(c)] and the lower-energy shift is observed at all measured momenta [Fig.~\ref{Fig.4}(d)].
In Fig.~\ref{Fig.4}(c), we also plot the charge gap ($2\Delta_{\rm C}$) evaluated in a study of optical conductivity~\cite{LSNOopticalII}, which approximately follows the energy shift of Charge I.
The ratio of the energy shift to the transition temperature of $\sim$7 is greater than that for conventional charge- and spin-density-wave transitions ($\sim$3.5~\cite{CDWSDW}), indicating that electron correlation plays a crucial role for the charge order as proposed from the temperature dependence of $2\Delta_{\rm C}$.
The fitting analysis shows that the energy of Charge II also decreases with increasing temperature.
In addition, peak width of Charge II broadens significantly. 
Since Charge II is the excitation associated with a spin-flip, spin stripe ordering may affect the lifetime of the excitation.

Our O $K$-edge RIXS results reveal that LSNO with $n_{h}$ = 0.33 has momentum-independent and temperature-dependent charge excitation at 0.6 eV, which originates from the doped holes, whereas in LSCO charge excitation at 0.5 eV has considerable dispersion but little temperature dependence~\cite{LSCOoxyRIXS}.
According to the optical conductivity studies, the temperature-dependent charge excitations are observed at 0.5 eV for LSNO~\cite{LSNOopticalII,RSNOoxyXAS,NSNOCO,LSNOopticalIV} while the variation with temperature is limited below 0.04 eV for La$_{1.875}$Ba$_{0.125}$CuO$_{4}$~\cite{LBCO}.
Therefore, the energy scale of the charge order differs between nickelates and cuprates.
The origin could be the polaron formation due to strong EP coupling.
The binding energies of polarons, corresponding to their size, were experimentally estimated from the charge excitation based on the optical conductivity, revealing a larger energy for LSNO than for LSCO~\cite{EPcouplingI}.
This is consistent with the prediction from local-density methods~\cite{EPcouplingIII}.
The temperature-dependent charge excitation observed in O $K$-edge RIXS indicates evolution of the temperature of localized polarons.
The magnitude of the polarons' binding energy could cause a greater enhancement of the charge-ordering transition temperature in nickelates than in cuprates.
In recent high-resolution RIXS studies on cuprates \cite{CDWRIXSI,CDWRIXSII,CDWRIXSIII}, charge excitations observed below 0.1 eV are discussed in relation with the charge order.
This also implies that energy scale of the charge order is below 0.1 eV in cuprates.

In summary, we performed O $K$-edge RIXS measurements for LSNO with $n_{h}$ = 0.10 and 0.33.
The charge excitations originating from doped holes were successfully observed.  
The results demonstrate that O $K$-edge RIXS is an effective technique for investigating the charge excitations in hole-doped charge-transfer insulators.
The observed charge excitations are independent of the momentum transfer, indicating the localized character of the doped holes, which contrasts with the delocalized character of holes in LSCO.
The localized charge excitations could be formed as a result of the stronger electron--phonon coupling and the weaker quantum spin fluctuation in LSNO.
Our O $K$-edge RIXS study reveals the different natures of charge excitations due to doped holes in LSNO and LSCO, which indicates a close connection between charge delocalization and metallicity.

We thank K. Yamazoe and U. J. Ralph for supporting the experiments and K. Takubo for fruitful discussions.
The resonant inelastic X-ray scattering measurements at SPring-8 BL07LSU were carried out as joint research by the Synchrotron Radiation Research Organization and the Institute for Solid State Physics, The University of Tokyo (Proposal No. 2018A7558).
This work was performed under the Inter-University Cooperative Research Program of the Institute for Materials Research, Tohoku University (Proposal Nos. 16K0005, 17K0067, and 18K0064).
This work was partially supported by JSPS KAKENHI (Grant No. 16H04004, and 19H05824), Japan.
K. Yamagami would like to acknowledge the support from the Motizuki Fund of the Yukawa Memorial Foundation.

\end{document}